\documentclass[11pt]{article}
\usepackage{vietnam,epsfig}

\def\Journal#1#2#3#4{{#1} {\bf #2}, #3 (#4)}

\def\PRL{\em Phys. Rev. Lett.}
\def\PRD{{\em Phys. Rev.} D}

\def\APJ{\em Astrophys. J.}
\def\APJS{\em Astrophys. J. Supp.}
\def\MNRAS{\em Mon. Not. R. Astron. Soc.}

\def\mco{\multicolumn}

\def\be{\begin{equation}}
\def\ee{\end{equation}}
\def\bea{\begin{eqnarray}}
\def\eea{\end{eqnarray}}

\begin{document}
\vspace*{4cm}
\title{FROM COBE TO WMAP: A DECADE OF DATA UNDER SCRUTINY}

\author{ Louise~M. Ord }

\address{Department of Astrophysics \& Optics, School of Physics, University of New South Wales, Sydney, NSW 2052, Australia}

\maketitle\abstracts{This talk, presented at the 5th Rencontres du
Vietnam 2004, gives a review of the cosmological implications of the
cosmic microwave background (CMB) data. The observational progress
that has been made over the past decade is discussed and recent
constraints on cosmological parameters are given.  The concordance
model is presented and the implications of the most recent data for
cosmological inflation, reionisation and dark energy are
described. Finally, the expectations for the next decade of CMB
cosmology are summarised.}

\section{Introduction}
On 24 April 1992, scientists announced that the Differential Microwave
Radiometer instrument aboard the COsmic Background Explorer satellite
(COBE-DMR) measured slight variations in the temperature of the cosmic
microwave background (CMB) of one part in 10$^5$ on scales of about
10$^{\circ}$~\cite{smoot92}.  Since the original COBE-DMR detection,
numerous ground based, balloon borne and satellite measurements of the
temperature fluctuations have been made at many different locations on
the sky. These CMB anisotropy measurements have brought us to an era
of precision cosmology, enabling us to directly probe the thermal
history of our universe.

There are a number of physical processes that produce anisotropies in
the microwave sky on a range of scales.  Primary fluctuations are
imprinted on the CMB during decoupling or even before.  For example,
acoustic oscillations are generated before decoupling through the
gravitational collapse of matter into the potential wells. Secondary
perturbations are generated between decoupling and the present.  These
include the distortion of the blackbody spectrum as photons pass
through regions of hot ionised gas. Refer to example reviews for more
details of the physics of the CMB anisotropies~\cite{hu97,cha04}.

The CMB harbours the imprint of the initial perturbations that seeded
the gravitational collapse of structure.  This signal has been
filtered through the dynamical evolution of the expanding universe.
The analysis of the CMB anisotropy data can therefore not only advance
our understanding of the process of structure formation but also
reveal fundamental properties of the universe such as its expansion
rate, geometry and the amount of matter it contains.  As the
anisotropy detections become increasingly accurate, we can test the
predictions of theoretical cosmology and extend our understanding of
our evolving universe.

In this talk, presented at the 5th Rencontres du Vietnam 2004, we give
an overview of the cosmological implications of the CMB data. We
briefly discuss the observational progress that has been made over the
past decade and comment on recent cosmological parameter constraints.
We describe the current concordant model and explain how
simultaneously analysing combinations of independent observational
data sets can tighten cosmological constraints. We also discuss the
implications of the current data for inflation, reionisation and dark
energy and summarise what we can expect the next decade to hold for
CMB cosmology.

\section{The Observational Data}
\subsection{First Light}
\begin{figure}
\begin{center}
\psfig{figure=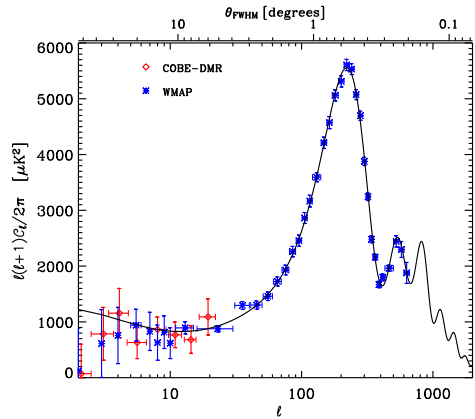,height=2.5in}
\caption{The power spectrum of temperature fluctuations from COBE-DMR
and WMAP. Measurements of the full sky enable the largest scales
(smallest multipoles) to be probed. Due to its small beam size ($\sim
21'$ FWHM), WMAP is also able to probe the smaller scales.
\label{f:fig1}}
\end{center}
\end{figure}

The potential for the CMB to become a potent probe of cosmology began
with the discovery of the tiny variations in the temperature of the
CMB by COBE-DMR in 1992.  Over a period of 4 years, COBE-DMR took
measurements of the full sky producing maps of the temperature
fluctuations at 3 frequencies. The dipole contribution was removed and
the galaxy was cut from the data to reveal a map of the CMB anisotropy
signal.  Variations on the spherical map occur on physical scales that
can be expanded in spherical harmonics to get a power spectrum of
temperature fluctuations (see Fig.~\ref{f:fig1}).  This is analogous
to Fourier expanding a flat map.

COBE-DMR's large beam size meant it was not able to probe small scales
(large multipoles).  However, because it made observations over the
whole sky, it was able to probe the largest scales (smallest
multipoles). These scales are larger than the physical horizon at the
time of recombination, so perturbations on these scales have not been
effected by interactions that occur during and after decoupling. These
perturbations therefore correspond to the initial density fluctuations
that have seeded structure formation.

A CMB photon that was last scattered from an over dense region will
have to overcome the gravitational attraction of that region in order
to make its way towards us.  As it does so, its wavelength is
stretched, so over dense regions correspond to cold spots on the CMB
map.  Similarly, the wavelength of a photon originating from an under
dense will be compressed as it makes its way towards us, so hot spots
correspond to under-densities.  This physics was first described in a
1967 paper by Sachs and Wolfe~\cite{sw67}.  The region probed by
COBE-DMR is therefore known as the Sachs-Wolfe Plateau.

\subsection{Unprecedented Precision}
\begin{figure} \begin{center}
\psfig{figure=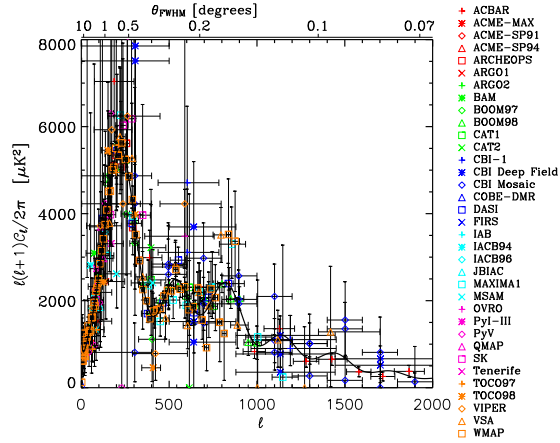,height=2.6in}
\caption{The current compilation of CMB data. The model plotted is the
concordant cosmological model: $\Omega_\Lambda \sim 0.72$, $\omega_b
\sim 0.0226$, $\omega_c \sim 0.123$, $n_{\rm s} \sim 0.96$, $\tau \sim
0.12$ and $h \sim 0.72$.
\label{f:fig2}}
\end{center} \end{figure}
\begin{figure} \begin{center}
\psfig{figure=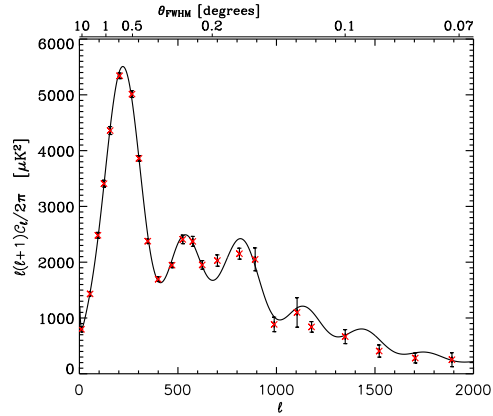,height=2.5in}
\caption{The current compilation of CMB data binned to better
visualise the fit of the concordance model.
\label{f:fig3}}
\end{center} \end{figure}
The most recent CMB data set to hit the headlines is WMAP, the
Wilkinson Microwave Anisotropy Probe~\cite{hin03}.  Scheduled to
observe for 4 years, the first year results were announced in February
2003.  Like COBE-DMR, WMAP is a differential experiment, measuring the
temperature difference between two points in the sky rather than
measuring absolute temperatures. The team combined information from 5
frequency channels and filtered out the signal from our galaxy in
order to get a full sky map.  You can see far more intricate detail
when comparing this map to the one from COBE-DMR.

The WMAP observations have 45 times the sensitivity and 33 times the
angular resolution of the COBE-DMR mission.  Again, as it's a full sky
map so probes the largest scales, the Sachs-Wolfe Plateau, but due to
its small beam size ($\sim 21'$ FWHM), it is also able to probe smaller
scales and the intricate physics that occurs during recombination and
at later times.  Fig.~\ref{f:fig2} shows the angular power spectrum
of temperature fluctuations from WMAP in its first year of
operation. The first year data sweeps over the first 2 peaks with
unprecedented accuracy.  These peaks correspond to acoustic
oscillations generated during recombination.

From COBE-DMR's first measurement of the CMB anisotropy to WMAP's most
recent observations there have been hundreds of detections made over a
range of scales by over 2 dozen autonomous and semi-autonomous
groups. Fig.~\ref{f:fig2} shows the current compilation of CMB data.
This data can be used to gain information about our universe such as
its matter content, its age and its geometry. It is difficult to
visualise the fit of the cosmological model with so much scatter, so
it is useful to bin the data (see Fig.~\ref{f:fig3}).  The model
plotted in Figs.~\ref{f:fig2} and~\ref{f:fig3} is the concordant
cosmological model.

\subsection{The Concordant Cosmological Model}
Although the CMB has the potential to simultaneously constrain a
number of cosmological parameters that are the ingredients of the hot
big bang model, particular parameter combinations can produce the same
spectra.  These model degeneracies limit our ability to extract
parameters from the CMB alone. Combining information from a range of
independent observational data sets enables certain degeneracies of
the individual data sets to be resolved.

\begin{table}
\caption{1$\sigma$ cosmological parameter constraints on flat
cosmologies from three analyses.\label{t:tab1}}
\vspace{0.4cm}
\begin{center}
\begin{tabular}{|c|c|c|c|c|c|}
\hline & \mco{2}{|c|}{Wang et al. (2002)~\cite{wan02}} &
\mco{2}{|c|}{Spergel et al. (2003)~\cite{spe03}} & Tegmark et
al. (2004)~\cite{teg04} \\

& CMB   & +2dFGRS & CMB(TT+TE) & +2dFGRS+Ly$\alpha$& CMB(TT+TE)+SDSS \\
\hline
$\Omega_\Lambda$  &$0.71 \pm 0.11$   & $0.72 \pm 0.09$  &$0.76^{+.05}_{-.06}$   &$0.74^{+.03}_{-.04}$   &$0.69^{+.03}_{-.04}$     \\
$\omega_b$        & 0.023$\pm$.003   & 0.024$\pm$.003   &$0.023\pm .001$        &$0.0226 \pm .0008$     &$0.0228^{+.0010}_{-.008}$\\
$\omega_c$        & 0.112$\pm$.014   & 0.115$\pm$.013   &$0.11^{+.06}_{-.04}$   &$0.11\pm .03$          &$0.123^{+.008}_{-.007}$   \\
$n_s$             &0.99$\pm$.06      & 0.99$\pm$.04     &$0.97 \pm .03$         &$0.96 \pm .02 $        &$0.96^{+.03}_{-.02}$      \\
$\tau$            & $0.04^{+0.06}$   &$0.06 \pm .03$    &$0.14^{+.07}_{-.06}$   &$0.12^{+.06}_{-.05}$   &$0.10^{+.06}_{-.05}$      \\
$h$               & 0.71$\pm$.13     & 0.73$\pm$.11     &$0.73 \pm .05$         &$0.72 \pm .03$         &$0.69 \pm .03$             \\
\hline
\end{tabular}
\end{center}
\end{table}
We are now beginning to refine an observationally concordant model; a
model that is concordant to a number of data sets.  Table~\ref{t:tab1}
gives the results of a few of the most recent analyses.  The $\chi^2$
per degree of freedom for the fit of the concordant model
($\Omega_\Lambda \sim 0.72$, $\omega_b \sim 0.0226$, $\omega_c \sim
0.123$, $n_{\rm s} \sim 0.96$, $\tau \sim 0.12$, $h \sim 0.72$) to the
full CMB anisotropy data set is $1.05$ with a 72\% probabilty of
finding a model that better fits the data \cite{gri04}, indicating
that the model provides a good fit to the anisotropy data alone.

\section{Implications for Inflation}
A key component of modern cosmologies is the assumption that the
initial irregularities are generated by a period of inflation during
the very early universe.  The simplest inflationary models predict an
approximately spatially flat universe, an approximately scale-invariant
spectrum of Gaussian, adiabatic primordial density perturbations and
an approximately scale-invariant spectrum of gravitational waves with
an amplitude that is a direct measure of Hubble rate during inflation.

\subsection{Spatial Flatness}
Flatness is probed through the positions of the acoustic peaks. Before
decoupling, the collapse of irregularities under gravity generates a
radiation pressure gradient.  This pressure gradient opposes the
gravitational collapse, causing the primordial plasma to acoustically
oscillate.  The phase of the acoustic oscillation is imprinted on the
photons as they last scatter producing the acoustic peaks in the
angular power spectrum~\cite{hu95,hu97}. Odd peaks represent phases
when the fluid in the potential wells is under compression and even
peaks rarefaction phases.

The phase of the oscillation, and thus the positions of the acoustic
peaks, depends on the maximum distance a sound wave could have
travelled since the Big Bang - the sound horizon. The curvature of the
universe affects the angle $\theta_{\rm s}$ subtended today by the
sound horizon at recombination.  If the subsequent journey of the CMB
photons was not distorted by large-scale cosmic curvature,
$\theta_{\rm s}\sim 1^{\circ}$. The position of the first peak tells
us that the largest CMB hot spots subtend an angle of about
$1^{\circ}$ on the sky. In fact, the locations of the CMB acoustic
peaks alone imply that the universe is approximately spatially flat
and there is no improvement on this prediction with the introduction
of other data.

\subsection{Adiabatic Initial Conditions}
The two general classes of perturbations, adiabatic and isocurvature,
drive acoustic oscillations that are out of phase from each other.
Adiabatic perturbations produce cosine harmonics and isocurvature
sine. They can therefore be distinguished by taking the ratios of the
locations of the acoustic peaks~\cite{hu96}.

Analyses have shown that adiabatic models provide a good fit to the
CMB data~\cite{spe03,teg04}.  Although there is no evidence for
isocurvature in the CMB, constraints are weak if correlated modes are
allowed~\cite{buc04}.  However, several cosmological parameters for
isocurvature models most favoured by the CMB conflict with other
probes, e.g. the baryonic density $\Omega_{\rm b}$ is much higher than
that inferred by light element abundances~\cite{buc04}.

\subsection{Gaussian Perturbations}
Measurements of the amplitude of non-Gaussian primordial fluctuations
are parameterised by a nonlinear coupling parameter, $f_{\rm NL}$. The
simplest inflationary models predict $\left|f_{\rm NL}\right|= 10^{-2}
- 10^{-1}$. The WMAP team tested their data for non-Gaussianity in two
ways~\cite{kom03}.  Firstly, constraints on $f_{\rm NL}$ using the
bispectrum gave $f_{\rm NL} = 38 \pm 48$ at 68\% confidence. They also
measured the morphology of the maps using Minkowski Functionals,
establishing the limit $f_{\rm NL} = 22 \pm 81$ at 68\% confidence.

The findings of the WMAP team reveal no evidence of primordial
non-Gaussianity, allowing them to conclude that their data is
consistent with Gaussian primordial perturbations. However, a number
of investigations have shown the WMAP data to harbour statistically
significant departures from rotational
invariance~\cite{oli04}$^-$\cite{muk04}.  Whether these effects are
primordial or systematic due to instrument effects or imperfect
foreground subtraction is yet unclear.

\subsection{Scale-Invariance}
Assuming a flat $\Lambda$CDM model with a power-law spectrum of
primordial scalar perturbations, the CMB alone implies a close to
scale-invariant spectrum in accordance with inflationary
predictions~\cite{spe03}. Slow roll inflation predicts a small
departure from a power-law spectrum, with a 2nd order running scalar
index: ${\rm d}n_{\rm s}/{\rm d} \ln k \sim (n_{\rm s}-1)^2$.  The
WMAP team reported weak evidence of running index by including
small-scale data from the 2-degree Field Galaxy Redshift Survey and
Ly$\alpha$ forest observations~\cite{spe03}. Their analysis implies a
much larger running than the slow roll prediction, although modelling
uncertainties in Ly$\alpha$ forest data questions the reliability of
this result~\cite{sel03}.

\begin{figure} \begin{center}
\psfig{figure=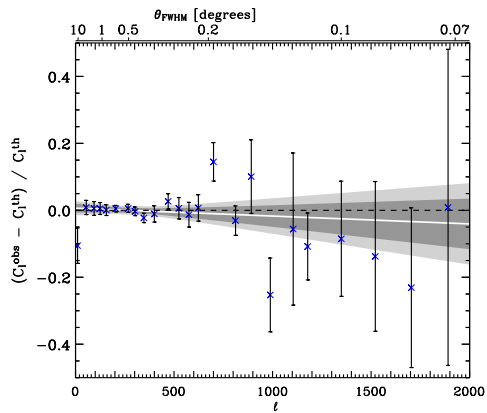,height=2.5in}
\caption{Residuals of the observed band power anisotropies with
respect to the concordant model plotted as a function of $\ell$. The
line that best-fits the data is plotted (solid white line) and the
68\% (dark grey region) and 95\% (light grey region) confidence
regions of the best-fit line are shaded. The concordant model is
consistent with the data although the high $\ell$ data prefers a model
with a slightly redder power-law spectrum, rendering the data on the
largest scales ($\ell<10$) anomalously low.
\label{f:fig4}}
\end{center} \end{figure}

The inclusion of the latest small-scale data such as VSA and CBI
reveals a preference for slightly redder power-law spectra,
e.g. $n_{\rm s}= 0.96^{+0.06}_{-0.03}$ combining VSA and
WMAP~\cite{reb04}. Furthermore, analyses that include this
interferometric data indicate a running in flat $\Lambda$CDM models at
the 2$\sigma$ level.  This reflects the fact that the data on the
largest scales ($\ell<10$) is anomalously low when tied to the latest
small scale data (see Fig.~\ref{f:fig4}~\cite{gri04}).  The evidence for a running
index is weakened considerably with inclusion of external priors from
large-scale structure data~\cite{teg04,reb04}. Future data on
both large and small scales will further test departures from a
scale-invariant primordial spectrum of scalar perturbations.

The generation of a nearly scale-invariant background of gravitational
waves is yet to be verified.  Current CMB limits on the
tensor-to-scalar ratio $r$ are weak, e.g. $r < 0.50$ at 95\%
confidence~\cite{teg04}.  However, interesting constraints in the
$r-n_{\rm s}$ plane for specific inflationary models are already being
made~\cite{teg04,lea03}.  The ultimate goal of future CMB
observations is to detect the B-mode polarisation signal that is
predicted to be generated by tensor mode perturbations from
gravitational waves.  Such detections would give a direct measurement
of the energy scale and place tight constraints on the dynamics of
inflation.

\section{Probing Reionisation and Dark Energy with the CMB}
\subsection{Polarisation}
Polarisation is generated by the final Thomson scattering of the
photons and therefore probes the redshift dependence of the optical
depth. Last scattering is expected to be concentrated in two distinct
redshift bands: $z \sim 1100$ (recombination) and below $z \sim 10$
(reionisation). In typical cosmological models the fraction of photons
that rescattered at low redshift ranges from a few percent to tens of
percent.

\begin{figure} \begin{center}
\psfig{figure=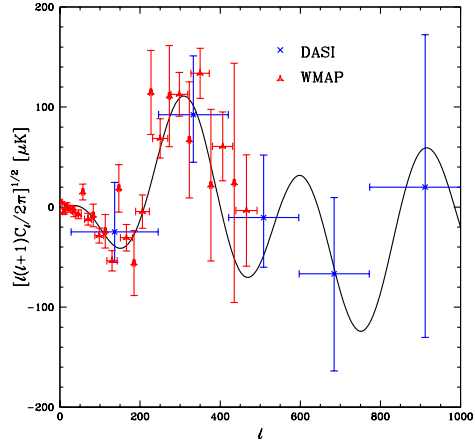,height=2.5in}
\caption{The T-E cross-correlation signal from DASI and WMAP. The
measurements on angular scales of about 1/2 a degree ($\ell \sim 300$)
provide evidence for super horizon size adiabatic fluctuations
existing on the surface of last scattering at recombination.
\label{f:fig5}}
\end{center} \end{figure}

Detection of the CMB polarisation signal was first announced by the
DASI team in 2002~\cite{kov02}. Assuming the concordant $\Lambda$CDM
model, they constrain both E- and B-mode spectra.  Their analysis
reveals a 5$\sigma$ detection of non-zero E-mode polarisation with an
amplitude consistent with the expected central value (see
Fig.~\ref{f:fig5}).  They report a T-E cross-correlation signal at
95\% significance but find no evidence for B-mode polarisation.

\begin{figure} \begin{center}
\psfig{figure=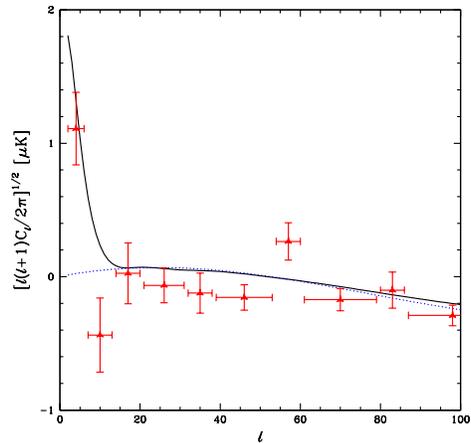,height=2.5in}
\caption{The WMAP data implies a significant excess power on large
scales ($\ell < 20$) over the prediction assuming polarisation is only
generated at recombination (dotted line). The solid line shows the
concordant model.
\label{f:fig6}}
\end{center} \end{figure}

The first year data release from the WMAP team includes detections of
the T-E cross correlation spectrum (see Fig.~\ref{f:fig5}). In
accordance with the inflationary prediction, the WMAP measurements on
angular scales of about 1/2 a degree ($\ell \sim 300$) provide
evidence for super horizon size adiabatic fluctuations existing on the
surface of last scattering at recombination.  They also find a
significant excess power on large scales ($\ell < 20$) over the
prediction assuming polarisation is only generated at recombination
(see Fig.~\ref{f:fig6}). Such an excess is consistent with an early
epoch of reionisation ($11 < z_{\rm re} < 30$, $\tau_{\rm re} = 0.17
\pm 0.04$) which is hard to reconcile with the significantly later
epoch expected from quasar absorption spectra observations. This
implies a complex reionisation history.

\subsection{Correlations with Gravity Tracers}
When the universe becomes dominated by dark energy the gravitational
potential decays, shifting the frequencies of photons passing though
it.  This local phenomenon generates the late-time integrated
Sachs-Wolfe effect that contributes to CMB power at the lowest
multipoles.  $\Lambda$CDM models should therefore produce large-angle
CMB anisotropies that are positively correlated with matter
fluctuations at $z < 2$, providing an independent check for dark
energy~\cite{cri96}.

The WMAP team~\cite{nol04} cross-correlated their data with the NVSS
radio source catalogue and found $\Lambda > 0$ with 95\% confidence
and $\Lambda = 0$ rejected at the 3$\sigma$ level. Boughn and
Crittenden~\cite{bou04} cross-correlated WMAP with data from the hard
X-ray background, which is dominated by active galaxy emission, and
the NVSS survey. They report positive correlations at 3$\sigma$ and
2.5$\sigma$ respectively.  Several groups have also detected the
cross-correlation on large scales between WMAP and optical galaxy
surveys~\cite{fos03,fos04}.

\section{Summary}
Precision cosmology has arrived. We are now able to probe the
ingredients of the hot big bang model with unprecedented accuracy.
Current ideas in cosmology appear to be vindicated. It seems we live
in a dark energy dominated universe in which more than 80\% of the
matter content is in the form of exotic cold dark matter. The CMB data
implies an almost spatially flat universe that evolved from an
approximately scale-invariant spectrum of Gaussian, adiabatic
primordial density perturbations. This is to be expected if quantum
fluctuations created in the very early universe were stretched to
astronomical scales during an epoch of accelerated expansion known as
cosmological inflation.

As we enter a new decade of CMB Observations, it is hoped that the
issues raised by the WMAP first-year data such as the possible
violation of rotational invariance and the implication of an early
epoch of reionisation are tested. Efforts continue to map the full CMB
sky to increasing sensitivity. The second-year data release of WMAP is
expected imminently and the Planck Satellite, due for launch in 2007,
intends to provide a full sky map of unprecedented accuracy and a
cosmic-variance limited measurement of the power spectrum up to $\ell
\sim 2000$.

We can also expect more data at high $\ell$ to high sensitivity. New
ground based experiments such as ACT, APEX and SPT are in progress.
Such small scale observations should provide further tests for
departures from a power-law primordial spectrum and begin to explore
secondary anisotropies such as the Sunyaev-Zeldovich and
Ostriker-Vishniac effects.

Efforts to constrain polarisation continue. BICEP, CAPMAP,
Polarbear, QUAD and SPORT are applying new techniques to produce
accurate measurements of E-mode polarisation and its temperature
anisotropy correlations and there is already talk of a post-Planck
satellite, CMBPOL, dedicated to CMB polarisation measurements. The
ultimate goal remains to detect B-mode signal generated by tensor mode
perturbations from gravitational waves.  Such detections would give a
direct measurement of the energy scale of inflation and place tight
constraints on the dynamics of inflation.

We can expect the CMB to continue to probe fundamental physics and
cosmology for years to come. So far it has provided a very good
determination of cosmological parameters but that isn't the same as
understanding why. Why dark energy and dark matter? Why inflation? Why
early reionisation? Now that the cosmological uncertainties have been
removed, the focus should turn to understanding the physics.



\section*{Acknowledgments}
It is a pleasure to thank Andrew Liddle and Charley Lineweaver for
interesting discussions and useful comments.  The author acknowledges
financial support from the Australian Research Council.


\section*{References}
\begin{thebibliography}{99}
\bibitem{smoot92} G.F. Smoot {\it et al.},
\Journal{\APJ}{396}{L1}{1992}.

\bibitem{hu97} W. Hu, N. Sugiyama, J. Silk, \Journal{\em
Nature}{386}{37}{1997}.

\bibitem{cha04} A. Challinor, {\em Proceedings of the 2nd Aegean
Summer School on the Early Universe}, 22-30 September 2003, Springer
LNP, astro-ph/0403344 (2004).

\bibitem{sw67} R.K. Sachs and A.M. Wolfe,
\Journal{\APJ}{147}{73}{1967}.

\bibitem{hin03} G. Hinshaw {\it et.al.},
\Journal{\APJS}{148}{135}{2003}.

\bibitem{wan02} X. Wang, M. Tegmark and M. Zaldarriaga,
\Journal{\PRD}{651}{123001}{2002}.

\bibitem{spe03} D.N. Spergel {\it et al},
\Journal{\APJS}{148}{175}{2003}.

\bibitem{teg04} M. Tegmark {\it et al},
\Journal{\PRD}{69}{103501}{2004}.

\bibitem{gri04}L.M. Griffiths and C.H. Lineweaver,
\Journal{\APJ}{603}{371}{2004}.

\bibitem{hu95} W. Hu, {\em ``Wandering in the Background: a Cosmic
Microwave Background Explorer''}, Ph.D. thesis, U. C. Berkeley (1995).

\bibitem{hu96} W. Hu, M. White, \Journal{\APJ}{471}{30}{1996}.

\bibitem{buc04} M. Bucher, J. Dunkley, P.G. Ferreira, K. Moodley and
C. Skordis, \Journal{\PRL}{93}{081301}{2004}.

\bibitem{kom03} E. Komatsu {\it et al},
\Journal{\APJS}{148}{119}{2003}.

\bibitem{oli04} A. de Oliveira-Costa, M. Tegmark, M. Zaldarriaga and
A. Hamilton, \Journal{\PRD}{69}{3516}{2004}.

\bibitem{vie04} P. Vielva, E. Mart\'{i}nez-Gonz\'{a}lez, R.B. Barreiro,
J.L. Sanz and L. Cay\'{o}n, \Journal{\APJ}{609}{22}{2004}.

\bibitem{cop04} C.J. Copi, D. Huterer and G.D. Starkman,
\Journal{\PRD}{70}{3515}{2004}.

\bibitem{eri04} H.K. Eriksen, D.I. Novikov, P.B. Lilje, A.J. Banday
and K.M. Gorski, \Journal{\APJ}{612}{64}{2004}.

\bibitem{han04} F.K. Hansen, P. Cabella, D. Marinucci and N. Vittorio,
\Journal{\APJ}{607}{L67}{2004}.

\bibitem{muk04} P. Mukherjee and Y. Wang, preprint astro-ph/0402602 (2004).

\bibitem{reb04} R. Rebolo {\it et al}, \Journal{\MNRAS}{353}{747}{2004}.

\bibitem{sel03} U. Seljak, P. McDonald and A. Makarov,
\Journal{\MNRAS}{342}{L79}{2003}.

\bibitem{lea03} S.M. Leach and A.R. Liddle,
\Journal{\MNRAS}{341}{1151}{2003}.

\bibitem{kov02} J.M. Kovac, E.M. Leitch, C. Pryke, J.E. Carlstrom,
N.W. Halverson and W.L. Holzapfel, \Journal{\em Nature}{420}{772}{2002}.

\bibitem{cri96} R.G. Crittenden and N. Turok,
\Journal{\PRL}{76}{575}{1996}.

\bibitem{nol04} M.R. Nolta {\it et al}, \Journal{\APJ}{608}{10}{2004}.

\bibitem{bou04} S. Boughn and R. Crittenden, \Journal{\em
Nature}{427}{45}{2004}.

\bibitem{fos03} P. Fosalba, E. Gazta\~{n}aga and F.J. Castander,
\Journal{\APJ}{597}{L89}{2003}.

\bibitem{fos04} P. Fosalba and E. Gazta\~{n}aga,
\Journal{\MNRAS}{350}{L37}{2004}.


\end{thebibliography}
\end{document}